\documentclass[authorversion,acmlarge]{acmart}

\acmYear{2020}\copyrightyear{2020}
\setcopyright{acmcopyright}
\acmConference[CryBlock 2020]{3rd Workshop on Cryptocurrencies and Blockchains for Distributed Systems}{September 25, 2020}{London, United Kingdom}
\acmBooktitle{3rd Workshop on Cryptocurrencies and Blockchains for Distributed Systems (CryBlock 2020), September 25, 2020, London, United Kingdom}
\acmPrice{15.00}
\acmDOI{10.1145/3410699.3413795}
\acmISBN{978-1-4503-8079-9/20/09}

\begin{document}

\title{Implications of Dissemination Strategies on the Security of Distributed Ledgers}

\author{Luca Serena}
\email{luca.serena2@unibo.it}
\affiliation{%
  \institution{CIRI ICT}
  \city{Cesena}
  \country{Italy}
}

\author{Gabriele D'Angelo}
\email{g.dangelo@unibo.it}
\affiliation{%
  \institution{University of Bologna}
  \city{Cesena}
  \country{Italy}
}

\author{Stefano Ferretti}
\email{stefano.ferretti@uniurb.it}
\affiliation{%
  \institution{University of Urbino}
  \city{Urbino}
  \country{Italy}
}


\begin{abstract}
This paper describes a simulation study on security attacks over Distributed Ledger Technologies (DLTs). We specifically focus on attacks at the underlying peer-to-peer layer of these systems, that is in charge of disseminating messages containing data and transaction to be spread among all participants. In particular, we consider the Sybil attack, according to which a malicious node creates many Sybils that drop messages coming from a specific attacked node, or even all messages from honest nodes. Our study shows that the selection of the specific dissemination protocol, as well as the amount of connections each peer has, have an influence on the resistance to this attack.
\end{abstract}




\maketitle

\section{Introduction}

Blockchain and Distributed Ledger Technologies (DLTs) have been widely employed and studied in the last years as a powerful and interesting tool to build novel decentralized applications that must provide some specific guarantees. Such guarantees are concerned with traceability, verifiability, data authenticity, and those applications requiring a tamper-proof data log \cite{ieeeaccess2020,sms2020,10.1145/3211933.3211950}. Being this technology seen as a secure building block for the construction of decentralized applications, most of the focus in research papers was on the application aspects \cite{ccnc2020}. Many security studies were devoted on the formal verification of certain tools for the provision of applications. For instance, a wide literature exists on the security aspects of smart contracts \cite{10.1145/3282373.3282419}. Other studies were devoted on comparative analyses of consensus algorithms, i.e.~the schemes that ensure that all the nodes sharing a copy of the ledger store a version that is eventually consistent with those maintained by others \cite{8400278,ZHANG202093}. Another aspect that received attention is on the network of transactions in a blockchain, whose analysis can reveal important insights for the de-anonymization of blockchain users \cite{8806723,sf-gda}.

Among the others, an aspect that received few attention was on mechanisms used for the lower level interaction of nodes that form the underlying peer-to-peer system. In fact, the typical approach is to consider the DLT as being built over some overlay, assuming it as a reliability network infrastructure where information is spread out. Yet, no focus was made on the security of this overlay, and on the impact that it might have on the blockchain (security) performance. This is not always the case  \cite{10.5555/2831143.2831152,10.1145/3327960.3332391}. For instance, the eclipse attacks consists in let an adversary controlling a sufficient number of IP addresses to monopolize all connections to and from a victim in the DLT peer-to-peer overlay.

In this paper, we study this possible threat by performed a specific simulation on the effects of Sybils attacks on DLTs. In particular, we consider an attack where a malicious node creates many Sybils in the peer-to-peer overlay, in order to obtain connections with honest nodes and obfuscate messages and transactions that need to be added in the ledger. We vary the topology of the overlay and the specific protocol, used to disseminate messages. Our results show that these varied metrics do have an influence on the success of the Sybil attack. Thus, they must be considered in order to build reliable and secure DLT infrastructures.

The remainder of this paper is organized as follows. Section~\ref{sec:back} introduces some background and related work. Section~\ref{sec:diss} provides discussion on the main dissemination protocols that can be used to spread messages in DLT networks. Section~\ref{sec:eval} describes the design of the experimental evaluation we conducted and the obtained results. Finally, Section~\ref{sec:conc} provides some concluding remarks.

\section{Background and Related Work}\label{sec:back}

In this section, we introduce some background and related work that is necessary for the rest of the paper. In particular, we briefly describe the distributed ledger technologies and the issues related to their simulation.

\subsection{Distributed Ledger Technologies (DLTs)}

A Distributed Ledger Technology (DLT) can be viewed as a database where there is no centralized data storage but information is distributed among a big part of the nodes of the system. These nodes have the same copy of the database, which can be read and edited independently by the single actors of the system.

Blockchain is probably the most famous example of distributed ledger. In this technology transactions are grouped into blocks, which are logically linked among each other. Blockchain-based systems use a consensus protocol to manage the flow of the blocks' validation. The blockchain can be treated as a protocol stack, in which each layer refers to a specific aspect of the blockchain. On the top of the stack, a consensus algorithm is used in order to let all nodes agree on the blockchain evolution.
Thus, the consensus scheme is in charge of ensuring that all nodes share the same view of of the shared ledger. Several schemes to reach a consensus exist, such as the Proof-of-Work, Proof of-Stake, Practical Byzantine Fault Tolerance \cite{8629877}.

The underlying layer consists of a peer-to-peer protocol. It is in charge of disseminating information on novel blocks being produced, to be added to the blockchain, or novel transactions that might be inserted into novel blocks. A flooding mechanism is often used to disseminate information, while the peer-to-peer overlay is built using some peer discovery mechanism. For instance, a random selection protocol is used in Bitcoin, while Ethereum employs a UDP-based node discovery mechanism inspired by Kademlia \cite{gencer2018}.

For which concerns the security, since in this kind of systems there is no server, centralized entity nor any kind of single point of failure, the assumptions to do in order to protect the system may differ. For example in a blockchain an attacker might try to break the consensus protocol, thus being able to control which transactions gets validated. An attacker might also want to isolate a certain user from the rest of the nodes, impeding him to receive and send any kind of data. We already mentioned the eclipse attack, according to which the attacker monopolizes the victim’s incoming and outgoing connections, thus isolating the victim from the other peers in the network \cite{10.5555/2831143.2831152}.

\subsection{Simulation of DLTs}

A strategy suitable to simulate a distributed ledger's behaviour is to use a time-stepped approach, in which the time is divided into discrete steps and all the nodes synchronize among each other before starting a new step. In each step of the simulation the simulated entities execute actions and receive messages (generating in turn other actions). Since in real life different actions may require a different duration of time to be executed, a multilevel approach can be used to improve the precision of the simulation. For example relaying a transaction requires much less time than mining a block in a proof-of-work system.

Only a few blockchain or DLT simulators are available in the state of the art. In \cite{aoki2019simblock}, an event-driven blockchain simulator is presented, that simulates the neighbor nodes selection of the peer-to-peer overlay. The mining activity is not simulated in detail, but a block generation is mimicked based on the computational capabilities of nodes.

The Bitcoin mining strategy is modeled and studied in \cite{gervais2016security}. In this work, only the network is modeled, but the propagation of transactions is not simulated. The focus was on the impact of the block size, block interval, and the block request management system.

VIBES is a blockchain simulator, thought for large modeling scale peer-to-peer networks \cite{vibes}. The design of the tool is thought to provide support for large-scale simulations with thousands of nodes.

BlockSim is a discrete-event simulator for blockchain systems \cite{10.1145/3308897.3308956}. BlockSim is organized in three layers: incentive layer, connector layer and system layer. Particular emphasis is given on the modeling and simulation of block creation through PoW.

Shadow-Bitcoin \cite{191667}, describes a methodology for the direct execution of multi-threaded applications inside a parallel discrete-event network simulation framework. This is used to implement a Bitcoin model.

In \cite{asiasim19}, we presented an agent-based simulator that reproduces some of the internals of a blockchain. This work is basically an extension of that preliminary work, based on a improved version of the LUNES-Blockchain simulator, more focused on security aspects of the dissemination protocol. In accordance to the multi-layered vision of a blockchain we previously discussed, a common approach is to simulate just few aspects of a blockchain at a time. Specifically, for the Sybil attack we consider in this work, it is not essential to represent every aspect of the blockchain, since we just want to evaluate the ability of a user chosen as victim to spread messages despite the presence of malicious nodes that don't forward them.

\section{Dissemination in DLTs}\label{sec:diss}

In this work, we specifically focus on the dissemination approaches that can be used to spread data (e.g.~transactions, blocks) in a DLT. The peer-to-peer overlay typically uses a decentralized configuration scheme, according to which every peer independently selects a subset of peers to connect with.

For instance, in Bitcoin, a node joining the network asks a set of special seed nodes (also known as Bitcoin DNS nodes) for a set of possible candidates to connect with \cite{8703385}. Then, the node may ask other peers, from this preliminary list, for
additional peers. This process can be iteratively repeated, in order to obtain a peer list. After retrieving such list, the node randomly selects a subset of peers and attempts to establish a connection, trying to maintain 8 active connections. In the meanwhile, the node can accept incoming connections from other peers, up to 117, that is the default maximum number of incoming connections.

As to message dissemination, there are several protocols that can be adopted. The simplest approach is to broadcast the received data to all the neighbors. To prevent infinite cycles of messages, two countermeasures can be used. The first one is the typical approach, in dissemination schemes, based on a caching system and a time to live, so as to drop old or already seen messages \cite{gda-jpdc-2017}. This mentioned approach relies on the use of a lower level caching system, independent from the data contained in the ledger data structures. However, it would be possible to check also data stored in the blockchain and in the node's mempool, to further reduce already seen information.

Broadcasting the messages is the strategy that grants the best coverage available but it's not efficient for minimizing the network traffic. In fact there are other protocols that allow either to reduce the number of messages sent or to enhance the anonymity of the transaction's creator. Some alternative algorithms that can be used for the dissemination of a transaction are \cite{gda-jpdc-2017}:
\begin{itemize}
\item \emph{Fixed Probability}: a message is sent to a neighbor only if a random generated number is greater than a certain threshold value. The operation is then repeated for every neighbor of the node, except for the forwarder of the message.
\item \emph{Probabilistic Broadcast}: a message is sent to all the neighbors (except the forwarder) only if a random generated number is greater than a certain threshold value. Otherwise no message is sent.
\item \emph{Dandelion}: a protocol aimed at enhancing the anonymity of the sender of a transaction \cite{10.1145/3084459}.
\end{itemize}
Dandelion is composed of two phases:
\begin{itemize}
\item Stem phase: the message is forwarded to a single neighbor, randomly chosen among the node's neighbors;
\item Fluff phase: here the broadcast of the message occurs, meaning that all the neighbors except the forwarder will receive the message.
\end{itemize}
An improved version of Dandelion exists, called Dandelion++ \cite{10.1145/3224424}, which aims to strengthen resilience against de-anonymization attacks and Sybil attacks. In particular, to avoid having transactions lost because of malicious or defective nodes that did not relay messages in the stem phase, the protocol implemented a fail-safe mechanism. If a node receives a message during the stem phase and it does not get it back after a certain amount of time in the fluff period, then that node will start the fluff phase, by broadcasting the message. Dandelion++ was recently adopted by the cryptocurrencies ZCoin \cite{zcoin} and Monero \cite{monero}, respectively in 2018 and in 2020.

Let's remember that in a blockchain system is not necessary a 100\% coverage. If a transaction doesn't reach a certain node then other nodes will mine it into a block and if a node misses to receive a block then a recovery function will be triggered in order to recover the information. However a high level of coverage is indispensable for the proper functioning of the system.

The performance of some these dissemination algorithms, in terms of network coverage and responsiveness, can be found in  previous research works such as \cite{gda-jpdc-2017}. However, as already mentioned, few attention was devoted to the choice of the proper dissemination algorithm in terms of security provision. In the next section, we will discuss on the performance of these considered dissemination strategies, also when varying the underlying topology of the peer-to-peer overlay.

\section{Performance Evaluation}\label{sec:eval}

In this section, we investigate the performance of different dissemination strategies, when varying the number of malicious nodes in the DLT and when varying the topology of the peer-to-peer overlay.

\subsection{Threat Model}

We consider a specific attack based on the presence of Sybils. The Sybil attack is a type of Denial of Service on peer-to-peer networks, in which the attacker takes multiple identities in order to enhance his influence in the system and carry out illegal actions \cite{10.1007/3-540-45748-8_24}. In this case, the attacker uses its Sybils to gain multiple connections with honest nodes, and thus influence the data they might receive, by filtering (incoming/outgoing) forwarded data. More specifically, in this study, we consider the case in which malicious nodes do not forward the transactions of a certain node, trying to impede him from spreading the transaction through the network.

\subsection{LUNES-Blockchain}

LUNES-Blockchain is a discrete event simulator which allows to simulate the behaviour of a blockchain running on top of a complex network topology. This paper is based on an enhanced version of the simulator that will freely available as source code on the research group website as part of the LUNES software distribution \cite{pads}. The simulator can be used for studying the normal behavior of the blockchain or in presence of attacks like 51\% attack, selfish mining or Sybil attack. When LUNES-Blockchain is used for studying malicious behaviors, the execution is repeated several times changing the computational power owned by the attacker (in the case of 51\% or selfish mining) or the number of malicious nodes in the system (in the case of the Sybil attack).

LUNES-blockchain is a modular simulator since it consists of three components that are executed separately and that can be easily replaced by other software modules:
\begin{itemize}
    \item \emph{Network creation}, managed by C library i-graph. It's possible to choose among different graph topologies (e.g. random graphs, k-regular graphs, small-world graphs, scale-free graphs) and to set the total number of nodes and edges in the network.
    \item \emph{Protocol simulation}. The actual execution is run, the user here can choose which attack to simulate and specific nodes behaviors.
    \item \emph{Performance evaluation}. Some scripts and tools are used to parse the data logged during the protocol simulation to assess the attacks outcomes and to generate statistical data.
\end{itemize}

\subsection{Setup and Methodology}

The performance evaluation reported in this paper considers random and small-world graphs with 10000 nodes. We used the mentioned topologies to mimic the structure of real world blockchains that have similar shapes. K-regular graphs were not taken into account because it's unlikely to have all that all nodes in the network have exactly the same number of outgoing edges. On the other hand, scale-free graphs have a hierarchical structure that is hard to find in actual distributed ledgers. For each graph configuration, the simulation was run 99 times, each time with an increasing percentage of malicious nodes in the network (i.e.~starting from 1 \% up to 99\%). The attackers (i.e.~Sybils) were chosen randomly among the nodes of the network. Thus, the number of honest nodes in the system decreases at each execution, proportionally with the increase of attackers.

Each simulation run consisted of 5000 discrete time steps and was divided into epochs, that are single tests executed over a different victim node. The final outcome is obtained by averaging the results of all the epochs run with a certain percentage of attackers. The number of epochs per execution is given by the ratio between the total number of time steps and the Time-To-Live (TTL) chosen for the delivered messages. The default TTL was set to 16, so in our setup there were 312 epochs in total. In the first step of an epoch, the victim sends a transaction and during the other steps the honest nodes forward the message containing the transaction. Conversely, the Sybils drop the message.

As discussed in the previous section, the scheme that it is used to create an overlay in a DLT system should maintain a certain randomness in the selection of nodes. For this reason, in the following we provide results obtained using two different topologies for the considered overlay networks, i.e.~random graphs and small worlds.
In future works, we will consider also other topologies, such as scale free networks, which are obtained when a preferential attachment is employed when adding nodes to the network \cite{gda-jpdc-2017}.

\subsection{Results}

\subsubsection{Random Graph Topology}

Firstly, we study the attack outcomes on a random undirected graph (built with Erdos-Renyi model) with 8 average outgoing edges per node.

\begin{figure}[t]
  \centering
    \includegraphics[width=0.8\textwidth]{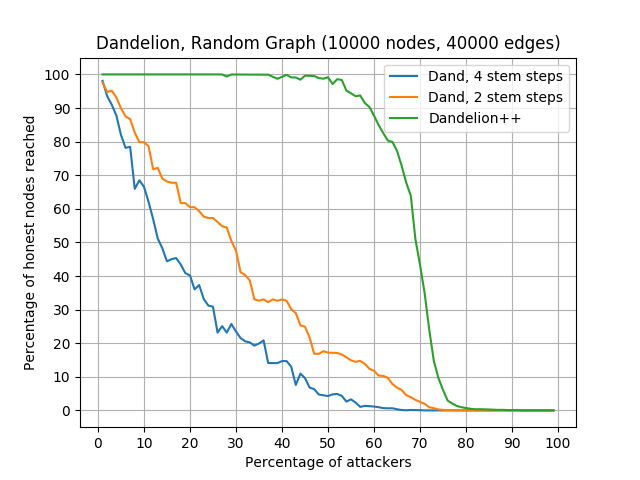}
  \caption{Coverage (Percentage of honest nodes reached) -- Dissemination: Dandelion, Overlay Topology: Random Graph with 40000 edges}
  \label{random-coverage-dandelion}
\end{figure}
\begin{figure}[t]
 \centering
\includegraphics[width=0.8\textwidth]{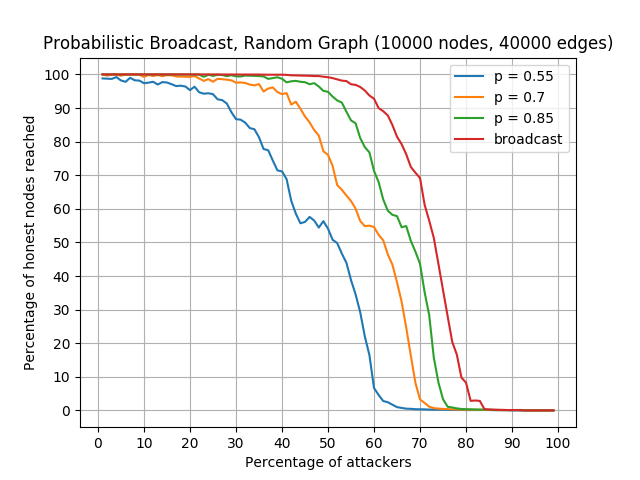}
  \caption{Coverage (Percentage of honest nodes reached) -- Dissemination: Probabilistic Broadcast, Overlay Topology: Random Graph with 40000 edges}
  \label{random-coverage-broadcast}
\end{figure}
\begin{figure}[t]
  \centering
\includegraphics[width=0.8\textwidth]{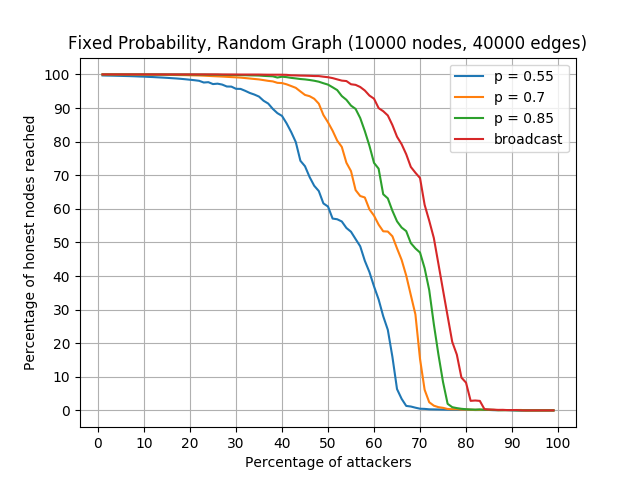}\\
  \caption{Coverage (Percentage of honest nodes reached) -- Dissemination: Fixed Probability, Overlay Topology: Random Graph with 40000 edges}
  \label{random-coverage-fixed}  
\end{figure}

As shown in Figure \ref{random-coverage-dandelion}, the coverage reached by Dandelion is lower than the other dissemination protocols (Figures \ref{random-coverage-broadcast}--\ref{random-coverage-fixed}). This means that bumping into a malicious node during the stem phase makes impossible the dissemination of the message. However, adopting the fail-safe mechanism provided by Dandelion++, the coverage can greatly improve. In this case, in fact, the protocol is able to reach performances similar to the Probabilistic Broadcast protocol (see Figure \ref{random-coverage-broadcast}). In fact Dandelion++ is actually very similar to pure broadcast in terms of coverage since if the stem phase is broken then the fail-safe mechanism will be activated, triggering the broadcast forwarding for nodes that have received the message during the stem phase (or the creator itself). However, in order to get similar outcomes, it is required to increase the TTL for messages at least as much as the number of steps a node waits before triggering, if necessary, the fail-safe mechanism. 
In these tests it was decided to make the nodes wait 6 discrete steps of time before activating the fail-safe mechanism. Note that without the presence of attackers, the number of steps necessary to get a full coverage is minor or equal than the maximum diameter of the graph for pure broadcast and a few more for other protocols. In our graphs, the maximum diameter is set to 10. However, having a big percentage of malicious nodes, the path to reach a node at the opposite end of the network can be much longer and winding. Therefore, in this condition, the messages (regardless of the protocol used) may run out of time before the maximum available coverage is reached. This problem is strongly minimized in graphs with a greater connectivity rate.

For what concerns Probabilistic Broadcast (see Figure \ref{random-coverage-broadcast}) and Fixed Probability (see Figure \ref{random-coverage-fixed}), it is necessary to specify that at the first step the creator of the transaction performs a full broadcast of the message. Otherwise, the network would lose a lot of messages during the first hop (in fact, if the conditional parameter's value is $n$, then there is $100-n\%$ of chance that the creator of a transaction is not sending any message). The performances between the two algorithms are very similar but Fixed Probability shows slightly better results.

\begin{figure}[t]
  \centering
\includegraphics[width=0.8\textwidth]{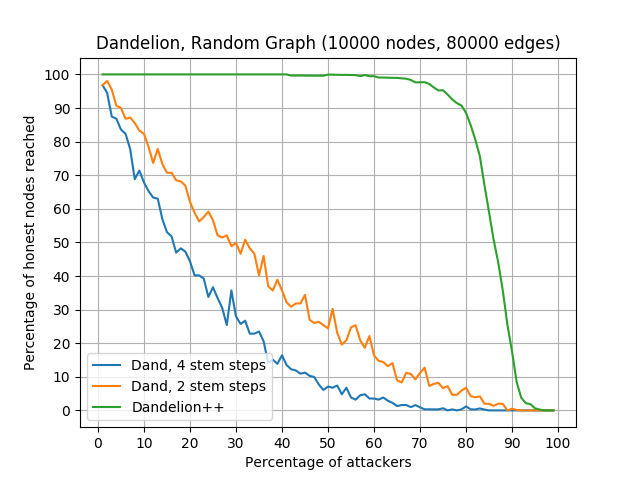}
  \caption{Coverage (Percentage of honest nodes reached) -- Dissemination: Dandelion, Overlay Topology: Random Graph with 80000 edges}
  \label{random-coverage-dandelion-80000}
\end{figure}

\begin{figure}[t]
  \centering
\includegraphics[width=0.8\textwidth]{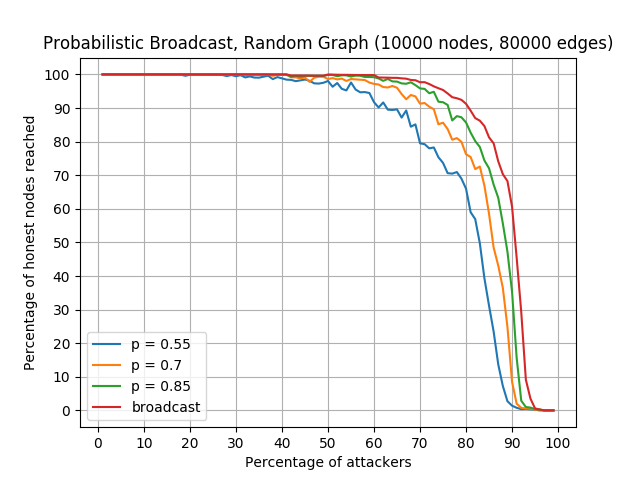}
  \caption{Coverage (Percentage of honest nodes reached) -- Dissemination: Probabilistic Broadcast, Overlay Topology: Random Graph with 80000 edges}
  \label{random-coverage-probabilistic-80000}  
\end{figure}

\begin{figure}[t]
  \centering
\includegraphics[width=0.8\textwidth]{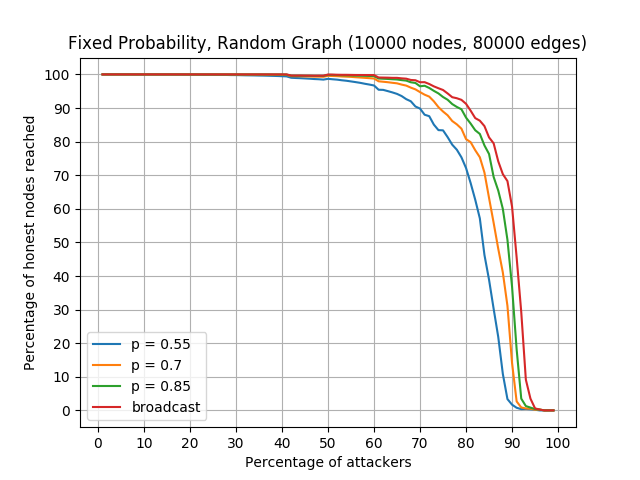}
  \caption{Coverage (Percentage of honest nodes reached) -- Dissemination: Fixed Probability, Overlay Topology: Random Graph with 80000 edges}
  \label{random-coverage-fixed-80000}
\end{figure}

In the following, we investigate how the number of connections in the network can influence the outcomes. As expected, with a larger number of connections, the attack is more difficult to carry out, since it's more probable for a node to have at least one honest neighbor to send the message to. In particular, all the protocols except Dandelion showed (see Figures \ref{random-coverage-dandelion-80000}, \ref{random-coverage-probabilistic-80000} and \ref{random-coverage-fixed-80000}) high resilience when the percentage of attackers is less than 70\%.


\subsubsection{Small-World Topology}

\begin{figure}[t]
  \centering
\includegraphics[width=0.8\textwidth]{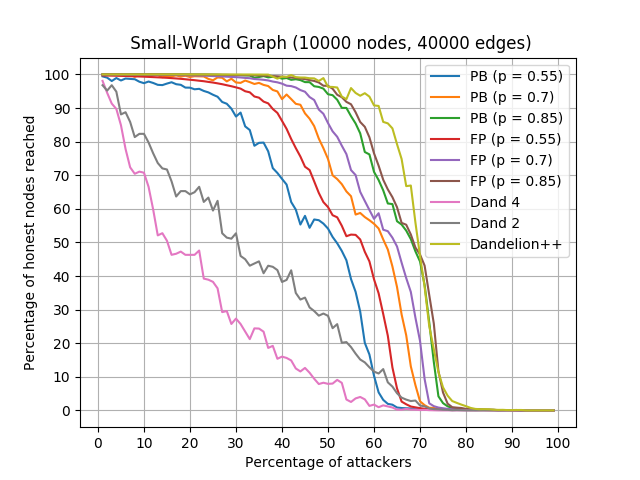}
  \caption{Coverage (Percentage of honest nodes reached) -- Dissemination: Dandelion, Fixed Probability (FP), Probabilistic Broadcast (PB), Overlay Topology: Small Worlds with 40000 edges}
  \label{small-coverage}  
\end{figure}

Having concluded that connectivity rate has a considerable influence on the attack outcome, let's figure out how the topology of the graph can affect the results. In Figure \ref{small-coverage} are reported the results obtained when running the simulation on a small-world graph with 8 average outgoing edges per node.

The outcomes reported in the figures are very similar to the simulation on Random Graphs with the same number of edges. In fact, all the observed patterns were repeated. Even in this case, the Fixed Probability protocol shows slightly better results than Probabilistic Broadcast and Dandelion behaves much worse than other protocols. 

To summarize, these findings are not sufficient to demonstrate that the network topology always has a negligible impact on the outcomes of the Sybil attack. In fact, some network topologies could make respectively easier or harder to complete the attack. However, we did not observe a significant difference between random graphs and the small worlds networks.

\section{Conclusions}\label{sec:conc}

In this paper, we performed a simulative study on the effects of Sybils attacks on DLTs. In particular, we considered an attack where a malicious node creates many Sybils in the peer-to-peer, in order to obtain connections with honest nodes and obfuscate messages and transactions that need to be added in the ledger.

It turns out that the number of connections in the network and the used dissemination protocol have an impact on the success of the attack. Thus, they are aspects to be considered in order to secure the communication infrastructure to build secure and reliable DLTs.
Dandelion is much more prone to Sybil attacks, due to its own inherent feature of relaying the message to just one node at the beginning. However, Dandelion's performance can be greatly improved adopting some Dandelion++ features. Dandelion++ should offer the coverage of pure broadcast and in addition assurances on sender's anonymity at the cost of more traffic of messages and a more complex management of transactions' relays. Another strategy to get a better level of coverage against a big number of attackers could be increasing the TTL of messages, since in presence of many malicious nodes, the path to connect two honest hosts may be much longer. This should not considerably increase the network traffic, because the caching system already allows to not forward to neighbors the already received messages. However, the biggest factor for resisting to Sybil attacks is the average number of connections among the nodes. Thus, having a highly interconnected network is the best guarantee for withstanding these malicious behaviours.

\bibliographystyle{ACM-Reference-Format}
\bibliography{biblio}


\begin{thebibliography}{27}


\ifx \showCODEN    \undefined \def \showCODEN     #1{\unskip}     \fi
\ifx \showDOI      \undefined \def \showDOI       #1{#1}\fi
\ifx \showISBNx    \undefined \def \showISBNx     #1{\unskip}     \fi
\ifx \showISBNxiii \undefined \def \showISBNxiii  #1{\unskip}     \fi
\ifx \showISSN     \undefined \def \showISSN      #1{\unskip}     \fi
\ifx \showLCCN     \undefined \def \showLCCN      #1{\unskip}     \fi
\ifx \shownote     \undefined \def \shownote      #1{#1}          \fi
\ifx \showarticletitle \undefined \def \showarticletitle #1{#1}   \fi
\ifx \showURL      \undefined \def \showURL       {\relax}        \fi
\providecommand\bibfield[2]{#2}
\providecommand\bibinfo[2]{#2}
\providecommand\natexlab[1]{#1}
\providecommand\showeprint[2][]{arXiv:#2}

\bibitem[\protect\citeauthoryear{??}{zco}{2018}]%
        {zcoin}
 \bibinfo{year}{2018}\natexlab{}.
\newblock \bibinfo{title}{Plus Plus Privacy: Zcoin Integrates Dandelion}.
\newblock
\newblock
\urldef\tempurl%
\url{https://zcoin.io/plus-plus-privacy-zcoin-integrates-dandelion/}
\showURL{%
\tempurl}


\bibitem[\protect\citeauthoryear{??}{mon}{2020}]%
        {monero}
 \bibinfo{year}{2020}\natexlab{}.
\newblock \bibinfo{title}{Another privacy-enhancing technology added to Monero:
  Dandelion++}.
\newblock
\newblock
\urldef\tempurl%
\url{https://web.getmonero.org/2020/04/18/dandelion-implemented.html}
\showURL{%
\tempurl}


\bibitem[\protect\citeauthoryear{Alharby and van Moorsel}{Alharby and van
  Moorsel}{2019}]%
        {10.1145/3308897.3308956}
\bibfield{author}{\bibinfo{person}{Maher Alharby} {and} \bibinfo{person}{Aad
  van Moorsel}.} \bibinfo{year}{2019}\natexlab{}.
\newblock \showarticletitle{BlockSim: A Simulation Framework for Blockchain
  Systems}.
\newblock \bibinfo{journal}{\emph{SIGMETRICS Perform. Eval. Rev.}}
  \bibinfo{volume}{46}, \bibinfo{number}{3} (\bibinfo{date}{Jan.}
  \bibinfo{year}{2019}), \bibinfo{pages}{135–138}.
\newblock
\showISSN{0163-5999}
\urldef\tempurl%
\url{https://doi.org/10.1145/3308897.3308956}
\showDOI{\tempurl}


\bibitem[\protect\citeauthoryear{Aoki, Otsuki, Kaneko, Banno, and Shudo}{Aoki
  et~al\mbox{.}}{2019}]%
        {aoki2019simblock}
\bibfield{author}{\bibinfo{person}{Yusuke Aoki}, \bibinfo{person}{Kai Otsuki},
  \bibinfo{person}{Takeshi Kaneko}, \bibinfo{person}{Ryohei Banno}, {and}
  \bibinfo{person}{Kazuyuki Shudo}.} \bibinfo{year}{2019}\natexlab{}.
\newblock \bibinfo{title}{SimBlock: A Blockchain Network Simulator}.
\newblock \bibinfo{howpublished}{arXiv:1901.09777}.
\newblock
\urldef\tempurl%
\url{https://arxiv.org/pdf/1901.09777.pdf}
\showURL{%
\tempurl}


\bibitem[\protect\citeauthoryear{{Bach}, {Mihaljevic}, and {Zagar}}{{Bach}
  et~al\mbox{.}}{2018}]%
        {8400278}
\bibfield{author}{\bibinfo{person}{L.~M. {Bach}}, \bibinfo{person}{B.
  {Mihaljevic}}, {and} \bibinfo{person}{M. {Zagar}}.}
  \bibinfo{year}{2018}\natexlab{}.
\newblock \showarticletitle{Comparative analysis of blockchain consensus
  algorithms}. In \bibinfo{booktitle}{\emph{2018 41st International Convention
  on Information and Communication Technology, Electronics and Microelectronics
  (MIPRO)}}. \bibinfo{pages}{1545--1550}.
\newblock


\bibitem[\protect\citeauthoryear{{Biryukov} and {Tikhomirov}}{{Biryukov} and
  {Tikhomirov}}{2019}]%
        {8806723}
\bibfield{author}{\bibinfo{person}{A. {Biryukov}} {and} \bibinfo{person}{S.
  {Tikhomirov}}.} \bibinfo{year}{2019}\natexlab{}.
\newblock \showarticletitle{Deanonymization and Linkability of Cryptocurrency
  Transactions Based on Network Analysis}. In \bibinfo{booktitle}{\emph{2019
  IEEE European Symposium on Security and Privacy (EuroS P)}}.
  \bibinfo{pages}{172--184}.
\newblock


\bibitem[\protect\citeauthoryear{Bojja~Venkatakrishnan, Fanti, and
  Viswanath}{Bojja~Venkatakrishnan et~al\mbox{.}}{2017}]%
        {10.1145/3084459}
\bibfield{author}{\bibinfo{person}{Shaileshh Bojja~Venkatakrishnan},
  \bibinfo{person}{Giulia Fanti}, {and} \bibinfo{person}{Pramod Viswanath}.}
  \bibinfo{year}{2017}\natexlab{}.
\newblock \showarticletitle{Dandelion: Redesigning the Bitcoin Network for
  Anonymity}.
\newblock \bibinfo{journal}{\emph{Proc. ACM Meas. Anal. Comput. Syst.}}
  \bibinfo{volume}{1}, \bibinfo{number}{1}, Article \bibinfo{articleno}{22}
  (\bibinfo{date}{June} \bibinfo{year}{2017}), \bibinfo{numpages}{34}~pages.
\newblock
\urldef\tempurl%
\url{https://doi.org/10.1145/3084459}
\showDOI{\tempurl}


\bibitem[\protect\citeauthoryear{{D'Angelo, Ferretti and Serena}}{{D'Angelo,
  Ferretti and Serena}}{2020}]%
        {pads}
\bibfield{author}{\bibinfo{person}{{D'Angelo, Ferretti and Serena}}.}
  \bibinfo{year}{2020}\natexlab{}.
\newblock \bibinfo{title}{PADS: Parallel and Distributed Simulation}.
\newblock
\newblock
\newblock
\shownote{Research Group, \url{http://pads.cs.unibo.it/}, accessed 16th June
  2020.}


\bibitem[\protect\citeauthoryear{Douceur}{Douceur}{2002}]%
        {10.1007/3-540-45748-8_24}
\bibfield{author}{\bibinfo{person}{John~R. Douceur}.}
  \bibinfo{year}{2002}\natexlab{}.
\newblock \showarticletitle{The Sybil Attack}. In
  \bibinfo{booktitle}{\emph{Peer-to-Peer Systems}},
  \bibfield{editor}{\bibinfo{person}{Peter Druschel}, \bibinfo{person}{Frans
  Kaashoek}, {and} \bibinfo{person}{Antony Rowstron}} (Eds.).
  \bibinfo{publisher}{Springer Berlin Heidelberg}, \bibinfo{address}{Berlin,
  Heidelberg}, \bibinfo{pages}{251--260}.
\newblock
\showISBNx{978-3-540-45748-0}


\bibitem[\protect\citeauthoryear{D’Angelo and Ferretti}{D’Angelo and
  Ferretti}{2017}]%
        {gda-jpdc-2017}
\bibfield{author}{\bibinfo{person}{Gabriele D’Angelo} {and}
  \bibinfo{person}{Stefano Ferretti}.} \bibinfo{year}{2017}\natexlab{}.
\newblock \showarticletitle{Highly intensive data dissemination in complex
  networks}.
\newblock \bibinfo{journal}{\emph{J. Parallel and Distrib. Comput.}}
  \bibinfo{volume}{99} (\bibinfo{year}{2017}), \bibinfo{pages}{28 -- 50}.
\newblock
\showISSN{0743-7315}
\urldef\tempurl%
\url{https://doi.org/10.1016/j.jpdc.2016.08.004}
\showDOI{\tempurl}


\bibitem[\protect\citeauthoryear{D’Angelo, Ferretti, and Marzolla}{D’Angelo
  et~al\mbox{.}}{2018}]%
        {10.1145/3211933.3211950}
\bibfield{author}{\bibinfo{person}{Gabriele D’Angelo},
  \bibinfo{person}{Stefano Ferretti}, {and} \bibinfo{person}{Moreno Marzolla}.}
  \bibinfo{year}{2018}\natexlab{}.
\newblock \showarticletitle{A Blockchain-Based Flight Data Recorder for Cloud
  Accountability}. In \bibinfo{booktitle}{\emph{Proceedings of the 1st Workshop
  on Cryptocurrencies and Blockchains for Distributed Systems}} (Munich,
  Germany) \emph{(\bibinfo{series}{CryBlock’18})}.
  \bibinfo{publisher}{Association for Computing Machinery},
  \bibinfo{address}{New York, NY, USA}, \bibinfo{pages}{93–98}.
\newblock
\showISBNx{9781450358385}
\urldef\tempurl%
\url{https://doi.org/10.1145/3211933.3211950}
\showDOI{\tempurl}


\bibitem[\protect\citeauthoryear{Fanti, Venkatakrishnan, Bakshi, Denby,
  Bhargava, Miller, and Viswanath}{Fanti et~al\mbox{.}}{2018}]%
        {10.1145/3224424}
\bibfield{author}{\bibinfo{person}{Giulia Fanti},
  \bibinfo{person}{Shaileshh~Bojja Venkatakrishnan}, \bibinfo{person}{Surya
  Bakshi}, \bibinfo{person}{Bradley Denby}, \bibinfo{person}{Shruti Bhargava},
  \bibinfo{person}{Andrew Miller}, {and} \bibinfo{person}{Pramod Viswanath}.}
  \bibinfo{year}{2018}\natexlab{}.
\newblock \showarticletitle{Dandelion++: Lightweight Cryptocurrency Networking
  with Formal Anonymity Guarantees}.
\newblock \bibinfo{journal}{\emph{Proc. ACM Meas. Anal. Comput. Syst.}}
  \bibinfo{volume}{2}, \bibinfo{number}{2}, Article \bibinfo{articleno}{29}
  (\bibinfo{date}{June} \bibinfo{year}{2018}), \bibinfo{numpages}{35}~pages.
\newblock
\urldef\tempurl%
\url{https://doi.org/10.1145/3224424}
\showDOI{\tempurl}


\bibitem[\protect\citeauthoryear{Ferretti and D'Angelo}{Ferretti and
  D'Angelo}{2020}]%
        {sf-gda}
\bibfield{author}{\bibinfo{person}{Stefano Ferretti} {and}
  \bibinfo{person}{Gabriele D'Angelo}.} \bibinfo{year}{2020}\natexlab{}.
\newblock \showarticletitle{On the Ethereum blockchain structure: A complex
  networks theory perspective}.
\newblock \bibinfo{journal}{\emph{Concurrency and Computation: Practice and
  Experience}} \bibinfo{volume}{32}, \bibinfo{number}{12}
  (\bibinfo{year}{2020}), \bibinfo{pages}{e5493}.
\newblock
\urldef\tempurl%
\url{https://doi.org/10.1002/cpe.5493}
\showDOI{\tempurl}
\newblock
\shownote{e5493 cpe.5493.}


\bibitem[\protect\citeauthoryear{Gencer, Basu, Eyal, Renesse, and Sirer}{Gencer
  et~al\mbox{.}}{2018}]%
        {gencer2018}
\bibfield{author}{\bibinfo{person}{Adem~Efe Gencer}, \bibinfo{person}{Soumya
  Basu}, \bibinfo{person}{Ittay Eyal}, \bibinfo{person}{Robbert Renesse}, {and}
  \bibinfo{person}{Emin~G{\"u}n Sirer}.} \bibinfo{year}{2018}\natexlab{}.
\newblock \showarticletitle{Decentralization in Bitcoin and Ethereum Networks}.
  In \bibinfo{booktitle}{\emph{Proceedings of the 22nd International Conference
  on Financial Cryptography and Data Security (FC). Springer}}.
\newblock
\urldef\tempurl%
\url{http://fc18.ifca.ai/preproceedings/75.pdf}
\showURL{%
\tempurl}


\bibitem[\protect\citeauthoryear{Gervais, Karame, W{\"u}st, Glykantzis, rf, and
  Capkun}{Gervais et~al\mbox{.}}{2016}]%
        {gervais2016security}
\bibfield{author}{\bibinfo{person}{Arthur Gervais}, \bibinfo{person}{Ghassan~O
  Karame}, \bibinfo{person}{Karl W{\"u}st}, \bibinfo{person}{Vasileios
  Glykantzis}, \bibinfo{person}{Hubert~Ritzdo rf}, {and}
  \bibinfo{person}{Srdjan Capkun}.} \bibinfo{year}{2016}\natexlab{}.
\newblock \showarticletitle{On the security and performance of proof of work
  blockchains}. In \bibinfo{booktitle}{\emph{Proceedings of the 2016 ACM
  SIGSAC}}. ACM, \bibinfo{pages}{3--16}.
\newblock


\bibitem[\protect\citeauthoryear{Heilman, Kendler, Zohar, and Goldberg}{Heilman
  et~al\mbox{.}}{2015}]%
        {10.5555/2831143.2831152}
\bibfield{author}{\bibinfo{person}{Ethan Heilman}, \bibinfo{person}{Alison
  Kendler}, \bibinfo{person}{Aviv Zohar}, {and} \bibinfo{person}{Sharon
  Goldberg}.} \bibinfo{year}{2015}\natexlab{}.
\newblock \showarticletitle{Eclipse Attacks on Bitcoin’s Peer-to-Peer
  Network}. In \bibinfo{booktitle}{\emph{Proceedings of the 24th USENIX
  Conference on Security Symposium}} (Washington, D.C.)
  \emph{(\bibinfo{series}{SEC’15})}. \bibinfo{publisher}{USENIX Association},
  \bibinfo{address}{USA}, \bibinfo{pages}{129–144}.
\newblock
\showISBNx{9781931971232}


\bibitem[\protect\citeauthoryear{Mense and Flatscher}{Mense and
  Flatscher}{2018}]%
        {10.1145/3282373.3282419}
\bibfield{author}{\bibinfo{person}{Alexander Mense} {and}
  \bibinfo{person}{Markus Flatscher}.} \bibinfo{year}{2018}\natexlab{}.
\newblock \showarticletitle{Security Vulnerabilities in Ethereum Smart
  Contracts}. In \bibinfo{booktitle}{\emph{Proceedings of the 20th
  International Conference on Information Integration and Web-Based
  Applications and Services}} (Yogyakarta, Indonesia)
  \emph{(\bibinfo{series}{iiWAS2018})}. \bibinfo{publisher}{Association for
  Computing Machinery}, \bibinfo{address}{New York, NY, USA},
  \bibinfo{pages}{375–380}.
\newblock
\showISBNx{9781450364799}
\urldef\tempurl%
\url{https://doi.org/10.1145/3282373.3282419}
\showDOI{\tempurl}


\bibitem[\protect\citeauthoryear{Miller and Jansen}{Miller and Jansen}{2015}]%
        {191667}
\bibfield{author}{\bibinfo{person}{Andrew Miller} {and} \bibinfo{person}{Rob
  Jansen}.} \bibinfo{year}{2015}\natexlab{}.
\newblock \showarticletitle{Shadow-Bitcoin: Scalable Simulation via Direct
  Execution of Multi-Threaded Applications}. In \bibinfo{booktitle}{\emph{8th
  Workshop on Cyber Security Experimentation and Test ({CSET} 15)}}.
  \bibinfo{publisher}{{USENIX} Association}, \bibinfo{address}{Washington,
  D.C.}
\newblock
\urldef\tempurl%
\url{https://www.usenix.org/conference/cset15/workshop-program/presentation/miller}
\showURL{%
\tempurl}


\bibitem[\protect\citeauthoryear{{Park}, {Im}, {Seol}, and {Paek}}{{Park}
  et~al\mbox{.}}{2019}]%
        {8703385}
\bibfield{author}{\bibinfo{person}{S. {Park}}, \bibinfo{person}{S. {Im}},
  \bibinfo{person}{Y. {Seol}}, {and} \bibinfo{person}{J. {Paek}}.}
  \bibinfo{year}{2019}\natexlab{}.
\newblock \showarticletitle{Nodes in the Bitcoin Network: Comparative
  Measurement Study and Survey}.
\newblock \bibinfo{journal}{\emph{IEEE Access}}  \bibinfo{volume}{7}
  (\bibinfo{year}{2019}), \bibinfo{pages}{57009--57022}.
\newblock


\bibitem[\protect\citeauthoryear{Rosa, D'Angelo, and Ferretti}{Rosa
  et~al\mbox{.}}{2019}]%
        {asiasim19}
\bibfield{author}{\bibinfo{person}{Edoardo Rosa}, \bibinfo{person}{Gabriele
  D'Angelo}, {and} \bibinfo{person}{Stefano Ferretti}.}
  \bibinfo{year}{2019}\natexlab{}.
\newblock \showarticletitle{Agent-Based Simulation of Blockchains}. In
  \bibinfo{booktitle}{\emph{Methods and Applications for Modeling and
  Simulation of Complex Systems}}, \bibfield{editor}{\bibinfo{person}{Gary
  Tan}, \bibinfo{person}{Axel Lehmann}, \bibinfo{person}{Yong~Meng Teo}, {and}
  \bibinfo{person}{Wentong Cai}} (Eds.). \bibinfo{publisher}{Springer
  Singapore}, \bibinfo{address}{Singapore}, \bibinfo{pages}{115--126}.
\newblock
\showISBNx{978-981-15-1078-6}


\bibitem[\protect\citeauthoryear{Stoykov, Zhang, and Jacobsen}{Stoykov
  et~al\mbox{.}}{2017}]%
        {vibes}
\bibfield{author}{\bibinfo{person}{Lyubomir Stoykov}, \bibinfo{person}{Kaiwen
  Zhang}, {and} \bibinfo{person}{Hans-Arno Jacobsen}.}
  \bibinfo{year}{2017}\natexlab{}.
\newblock \showarticletitle{VIBES: Fast Blockchain Simulations for Large-Scale
  Peer-to-Peer Networks: Demo}. In \bibinfo{booktitle}{\emph{Proceedings of the
  18th ACM/IFIP/USENIX Middleware Conference: Posters and Demos}} (Las Vegas,
  Nevada) \emph{(\bibinfo{series}{Middleware ’17})}.
  \bibinfo{publisher}{Association for Computing Machinery},
  \bibinfo{address}{New York, NY, USA}, \bibinfo{pages}{19–20}.
\newblock
\showISBNx{9781450352017}
\urldef\tempurl%
\url{https://doi.org/10.1145/3155016.3155020}
\showDOI{\tempurl}


\bibitem[\protect\citeauthoryear{{Wang}, {Hoang}, {Hu}, {Xiong}, {Niyato},
  {Wang}, {Wen}, and {Kim}}{{Wang} et~al\mbox{.}}{2019}]%
        {8629877}
\bibfield{author}{\bibinfo{person}{W. {Wang}}, \bibinfo{person}{D.~T. {Hoang}},
  \bibinfo{person}{P. {Hu}}, \bibinfo{person}{Z. {Xiong}}, \bibinfo{person}{D.
  {Niyato}}, \bibinfo{person}{P. {Wang}}, \bibinfo{person}{Y. {Wen}}, {and}
  \bibinfo{person}{D.~I. {Kim}}.} \bibinfo{year}{2019}\natexlab{}.
\newblock \showarticletitle{A Survey on Consensus Mechanisms and Mining
  Strategy Management in Blockchain Networks}.
\newblock \bibinfo{journal}{\emph{IEEE Access}}  \bibinfo{volume}{7}
  (\bibinfo{year}{2019}), \bibinfo{pages}{22328--22370}.
\newblock


\bibitem[\protect\citeauthoryear{Zhang and Lee}{Zhang and Lee}{2019}]%
        {10.1145/3327960.3332391}
\bibfield{author}{\bibinfo{person}{Shijie Zhang} {and}
  \bibinfo{person}{Jong-Hyouk Lee}.} \bibinfo{year}{2019}\natexlab{}.
\newblock \showarticletitle{Eclipse-Based Stake-Bleeding Attacks in PoS
  Blockchain Systems}. In \bibinfo{booktitle}{\emph{Proceedings of the 2019 ACM
  International Symposium on Blockchain and Secure Critical Infrastructure}}
  (Auckland, New Zealand) \emph{(\bibinfo{series}{BSCI ’19})}.
  \bibinfo{publisher}{Association for Computing Machinery},
  \bibinfo{address}{New York, NY, USA}, \bibinfo{pages}{67–72}.
\newblock
\showISBNx{9781450367868}
\urldef\tempurl%
\url{https://doi.org/10.1145/3327960.3332391}
\showDOI{\tempurl}


\bibitem[\protect\citeauthoryear{Zhang and Lee}{Zhang and Lee}{2020}]%
        {ZHANG202093}
\bibfield{author}{\bibinfo{person}{Shijie Zhang} {and}
  \bibinfo{person}{Jong-Hyouk Lee}.} \bibinfo{year}{2020}\natexlab{}.
\newblock \showarticletitle{Analysis of the main consensus protocols of
  blockchain}.
\newblock \bibinfo{journal}{\emph{ICT Express}} \bibinfo{volume}{6},
  \bibinfo{number}{2} (\bibinfo{year}{2020}), \bibinfo{pages}{93 -- 97}.
\newblock
\showISSN{2405-9595}
\urldef\tempurl%
\url{https://doi.org/10.1016/j.icte.2019.08.001}
\showDOI{\tempurl}


\bibitem[\protect\citeauthoryear{Zichichi, Ferretti, and D'Angelo}{Zichichi
  et~al\mbox{.}}{2020a}]%
        {ccnc2020}
\bibfield{author}{\bibinfo{person}{Mirko Zichichi}, \bibinfo{person}{Stefano
  Ferretti}, {and} \bibinfo{person}{Gabriele D'Angelo}.}
  \bibinfo{year}{2020}\natexlab{a}.
\newblock \showarticletitle{A Distributed Ledger Based Infrastructure for Smart
  Transportation System and Social Good}. In \bibinfo{booktitle}{\emph{IEEE
  Consumer Communications and Networking Conference (CCNC)}} (Las Vegas, USA).
\newblock


\bibitem[\protect\citeauthoryear{Zichichi, Ferretti, and D'Angelo}{Zichichi
  et~al\mbox{.}}{2020b}]%
        {ieeeaccess2020}
\bibfield{author}{\bibinfo{person}{Mirko Zichichi}, \bibinfo{person}{Stefano
  Ferretti}, {and} \bibinfo{person}{Gabriele D'Angelo}.}
  \bibinfo{year}{2020}\natexlab{b}.
\newblock \showarticletitle{A Framework based on Distributed Ledger
  Technologies for Data Management and Services in Intelligent Transportation
  Systems}.
\newblock \bibinfo{journal}{\emph{IEEE Access}} (\bibinfo{year}{2020}).
\newblock


\bibitem[\protect\citeauthoryear{Zichichi, Ferretti, and D'Angelo}{Zichichi
  et~al\mbox{.}}{2020c}]%
        {sms2020}
\bibfield{author}{\bibinfo{person}{Mirko Zichichi}, \bibinfo{person}{Stefano
  Ferretti}, {and} \bibinfo{person}{Gabriele D'Angelo}.}
  \bibinfo{year}{2020}\natexlab{c}.
\newblock \showarticletitle{On the Efficiency of Decentralized File Storage for
  Personal Information Management Systems}. In \bibinfo{booktitle}{\emph{Proc.
  of 2nd International Workshop on Social Media Sensing (SMS 2020) - 25th IEEE
  Symposium on Computers and Communications (ISCC)}}.
\newblock


\end{thebibliography}

\end{document}